\documentclass[prl,nofootinbib,twocolumn,showpacs,preprintnumbers,amsmath,amssymb,floatfix,superscriptaddress]{revtex4}
\usepackage{amsmath}
\usepackage{amsfonts}
\usepackage{dcolumn}                    	     	
\usepackage{amssymb}
\usepackage{graphicx,graphics,wrapfig,rotating}		
\usepackage[english]{babel}	
\usepackage{bm,fancybox}                	     	
\usepackage{bbm}

\usepackage{subfigure}

\def\bra#1{{\langle#1|}}
\def\ket#1{{|#1\rangle}}
\def\bracket#1#2{{\langle#1|#2\rangle}}

\begin{document}

\title{Stable classical structures in dissipative quantum chaotic systems}

\author{Lisandro A. Raviola}
\affiliation{Departamento de F\'\i sica, CNEA, Av. Libertador 8250, (C1429BNP) Buenos Aires, Argentina}
\author{Gabriel G. Carlo} 
\affiliation{Departamento de F\'\i sica, CNEA, Av. Libertador 8250, (C1429BNP) Buenos Aires, Argentina}
\author{Alejandro M. F. Rivas} 
\affiliation{Departamento de F\'\i sica, CNEA, Av. Libertador 8250, (C1429BNP) Buenos Aires, Argentina}
\affiliation{Instituto de Ciencias, Universidad Nacional de General Sarmiento, J.M. Gutierrez 1150, (B1613GSX) Los Polvorines, Buenos Aires, Argentina}

\date{\today}


\begin{abstract}

We study the stability of classical structures in chaotic systems when a 
dissipative quantum evolution takes place. We consider a paradigmatic model, 
the quantum baker map in contact with a heat bath at finite temperature. 
We analyze the behavior of the purity, fidelity and Husimi distributions corresponding 
to initial states localized on short periodic orbits (scar functions) and map eigenstates. 
Scar functions, that have a fundamental role in the semiclassical description of chaotic systems, 
emerge as very robust against environmental perturbations. 
This is confirmed by the study of other states localized on classical structures. 
Also, purity and fidelity show a complementary behavior as
decoherence measures. 

\end{abstract}

\maketitle


Since the origin of quantum theory until present the correspondence between the 
classical and quantum views of nature has been a major source of debate. 
Two broad areas of physics have converged naturally towards its study. On one 
hand we have the semiclassical theory of closed systems, where the problem consists of 
linking the Hamiltonian with the unitary quantum evolution as $\hbar \rightarrow 0$.
For integrable systems the EBK quantization scheme \cite{Brack} provides a 
precise correspondence between quantum numbers and energy levels and between 
eigenfunctions and invariant tori. For chaotic systems the correspondence is 
not so precise but the Gutzwiller trace formula \cite{Gut} and more recently, 
the short periodic orbits theory \cite{SPOth} are resources at our disposal. 
However, a complete description 
of chaotic eigenfunctions in terms of classical invariants is still lacking. On the 
other hand, we have the decoherence theory that has established a correspondence 
between dissipative quantum systems and their classical analogs \cite{ZurekPaz}. 
These studies have enormous relevance nowadays not only from the theoretical, but 
also from the experimental point of view. In fact, 
decoherence poses an unavoidable difficulty to 
the coherent manipulation of small scale systems, which in addition generally 
have a complex dynamics. Hence quantum mechanics, nonlinear dynamics and 
decoherence are necessary ingredients in many areas of physics, involved in a 
broad range of theoretical models and implementations. Just a few examples 
are transport in cold atoms and BECs \cite{CAandBECs}, nanodevices 
\cite{nanodevices}, microlasers \cite{Microlasers,Microlasers2,WiersigPrl06}, 
quantum dots \cite{QuantumDots}, and chaotic scattering 
\cite{ChaoticScattering}.

In view of this we can ask ourselves, are there any classical structures 
embedded in the quantum realm that emerge as specially resistant to external 
perturbations? If so, in which way? To answer this questions, 
in this letter we investigate the stability of initial states with different 
degrees of classical information under the effects of chaotic evolution and 
decoherence. The environment is introduced by coupling the system 
to a heat bath at finite temperature, in a way close to actual experimental situations. 
In order to capture all the essential features of dissipative quantum chaotic 
systems without unnecessary complications, we focus our study on dissipative 
quantum maps \cite{Bianucci}. They 
constitute an ideal testbed for semiclassical and decoherence theories. 

We find that scar functions, which are states localized along the 
stable and unstable manifolds of 
periodic orbits, are highly stable with respect to environmental perturbations. 
We quantify this stability by studying the purity and the fidelity, 
quantities that show complementary in order to measure decoherence.
Moreover, by analyzing the Husimi distributions in phase space we 
conclusively show this behavior. 
 

We perform the evolution of the density matrix of 
the system by means of a two step operator $\rho$ \cite{Bianucci,Garcia-Mata}. 
We use a composition of a unitary step given by the closed map $B$ (representing the 
system dynamics), and a purely dissipative step given by the 
superoperator ${\bf D}_{\alpha}$, in the form
\begin{equation}
 \rho' = {\bf S}(\rho) = {\bf D}_\alpha ({\bf B}(\rho) )
\label{S}
\end{equation}
In this equation ${\bf B}$ represents the unitary superoperator and 
$\alpha$ all the parameters of the environmental model.
In the following we explain the construction of these two steps.

We consider the baker map $\mathcal{B}$ on the unit 
torus \cite{Voros} as the system. It is given by $(q', p') = 
\mathcal{B}(q,p) = (2q-[2q], p+[2q]/2)$ where $(q,p)$ are the 
position and momentum coordinates and $[x]$ stands for the integer part of $x$. 
This transformation is an area-preserving, uniformly hyperbolic, 
piecewise-linear and invertible map with Lyapunov exponent $\lambda=\ln{2}$. 
The phase space has a very simple Markov partition consisting of 
two regions ($q<1/2$ and $q\geq1/2$) associated with the symbols $0$ and $1$, 
for which there is a complete symbolic dynamics. The action of the map upon 
symbols can be understood by means of the binary expansion of the coordinates
$(p|q)=\ldots\nu_{-1}\cdot\nu_{0} 
\nu_{1}\ldots
\stackrel{\mathcal{B}}{\longrightarrow}
(p\prime|q\prime)=\ldots\nu_{-1}\nu_{0}
\cdot\nu_{1}\ldots$
where $q=\sum_{i=0}^{\infty}\nu_{i} 2^{-(i+1)}$ and 
$p=\sum_{i=-1}^{-\infty}\nu_{i} 2^{i}$.
Then, a periodic orbit (PO) of period $L$ can be represented 
by a binary string $\bm{\nu}$ of length $L$.
The coordinates of the first trajectory point $(q_0,p_0)$ on the periodic orbit can be 
obtained explicitly in terms of the binary string as
$q_0=\cdot\bm{\nu\nu\nu}\ldots=\nu/(2^{L}-1)$ and 
$p_0=\cdot\bm{\nu^{\dagger}\nu^{\dagger}\nu^{\dagger}}
\ldots=\nu^\dagger/(2^{L}-1)$, where $\nu$ is the integer value of 
the string $\bm{\nu}$ which represents a binary number, and 
$\bm{\nu^{\dagger}}$ is the string formed by all $L$ bits of 
$\bm{\nu}$ in reverse order. The other trajectory points can be easily calculated by 
iterations of the map or by cyclic shifts of $\bm{\nu}$.

When quantizing this system any state $\ket{\psi}$ must satisfy periodic 
boundary conditions on the torus, for both the position and momentum representations. This 
amounts to taking $\bracket{q+1}{\psi}\:=\:e^{i 2 \pi \chi_q}\bracket{q}{\psi}$, and
$\bracket{p+1}{\psi}\:=\:e^{i 2 \pi \chi_p}\bracket{p}{\psi}$, 
with $\chi_q$, $\chi_p \in [0,1)$. This implies a 
Hilbert space of finite dimension $N=(2 \pi \hbar)^{-1}$. 
The discrete set of position and momentum eigenstates is given by
$\ket{q_j}\:=\:\ket{(j+\chi_q)/N}\;(j=0, 1,  \dots N-1)$, and
$\ket{p_k}\:=\:\ket{(k+\chi_p)/N}\;(k=0, 1,  \dots N-1)$, 
labeled by the corresponding eigenvalues $q_j$, $p_k$.
They are related by a discrete Fourier transform, i.e. $\bracket{p_k}{q_j}\:=\:
1/\sqrt{N} \exp (-i (2 \pi/N) (j+\chi_q)(k+\chi_p)) \: \equiv \: (G^{\chi_q, \chi_p}_N)$. 
Throughout the paper we assume a phase space with anti-symmetric boundary conditions 
($\chi_q=\chi_p=1/2$).
The unitary operator $\hat{B}$ that performs the closed quantum evolution 
is given by \cite{Voros,Saraceno}:
\begin{equation}
\hat{B}=G^\dagger_N
\left(\begin{array}{cc}
G_{N/2}& 0 \\       
0 & G_{N/2} \\
\end{array}\right).
\label{quantumbaker}
\end{equation}

We incorporate dissipation and thermalization to the quantum map by coupling it to 
a bath of noninteracting oscillators in thermal equilibrium at a temperature $T$. 
The degrees of freedom of the bath can be eliminated by means of the usual weak coupling, 
Markov and rotating wave approximations \cite{QNoise}. As a result we arrive at a 
Lindblad equation for the density matrix of the system $\rho$ that can 
be written as a completely positive map ${\bf D}_\alpha(dt)$ in the operator-sum 
(or Kraus) representation
\begin{equation}
\rho(t+dt) = {\bf D}_{(\varepsilon,T)}(dt)\left(\rho(t)\right)=\sum_{\mu=0}^{2}{K_\mu \rho(t) K^\dagger_\mu}
\label{krausrep}
\end{equation}
where 
\begin{eqnarray}
K_0 &=& \mathbbm{1}-\frac{1}{2} \; \sum_{\mu=1}^{2}{K^\dagger_\mu K_\mu} \nonumber \\
K_1 &=& \sum_{k=1}^{N-1} \sqrt{\varepsilon \;dt \; (1+\bar{n}(k)) \; k} \; \ket{p_{k-1}}\bra{p_{k}}  \nonumber \\
K_2 &=& \sum_{k=1}^{N-1} \sqrt{\varepsilon \;dt \; \bar{n}(k) \; k} \; \ket{p_{k}}\bra{p_{k-1}}
\label{krauslindblad}
\end{eqnarray}
are the infinitesimal Kraus operators satisfying $\sum_\mu{K^\dagger_\mu K_\mu}=\mathbbm{1}$ 
to first order in $dt$ \cite{Carlo}. In these equations $\varepsilon$ is a system-bath coupling parameter that 
can be associated to a classical velocity dependent damping (at $T=0$ gives the contraction rate 
of the phase space). The population densities of the bath are given by $\bar{n} = 
(\exp(\Delta E_{k}/(k_B T))-1)^{-1}$, where we have taken $E_{k} = p_{k}^2/2$, and the Boltzmann 
constant $k_B=1$. Then, we integrate ${\bf D}_{(\varepsilon,T)}(dt)$ numerically 
to obtain the dissipative step. 

\begin{figure}

 \subfigure[ POM] {
 \includegraphics[width=0.4\textwidth]{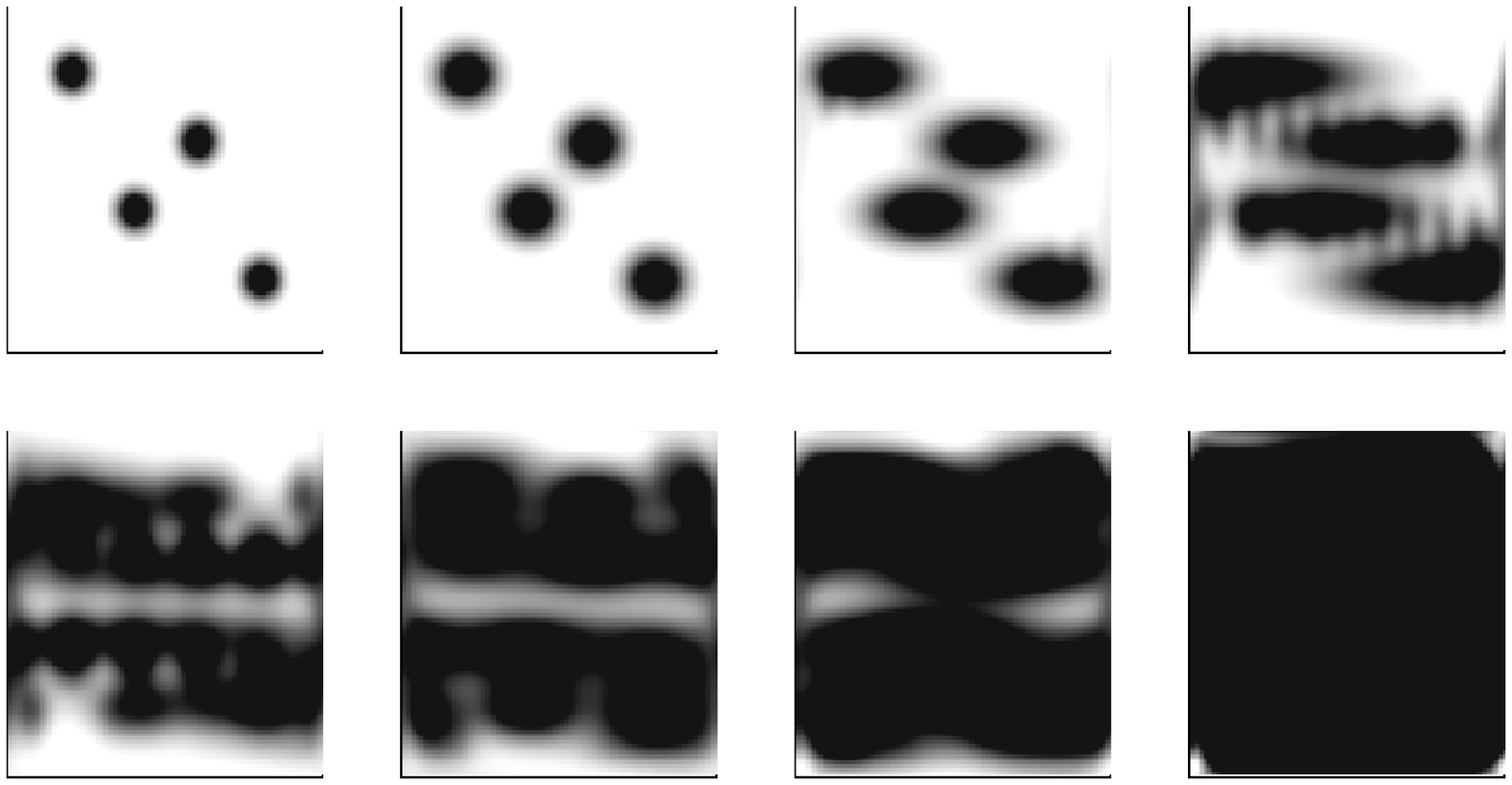} 
 }
 
 \subfigure[ Scar] {
 \includegraphics[width=0.4\textwidth]{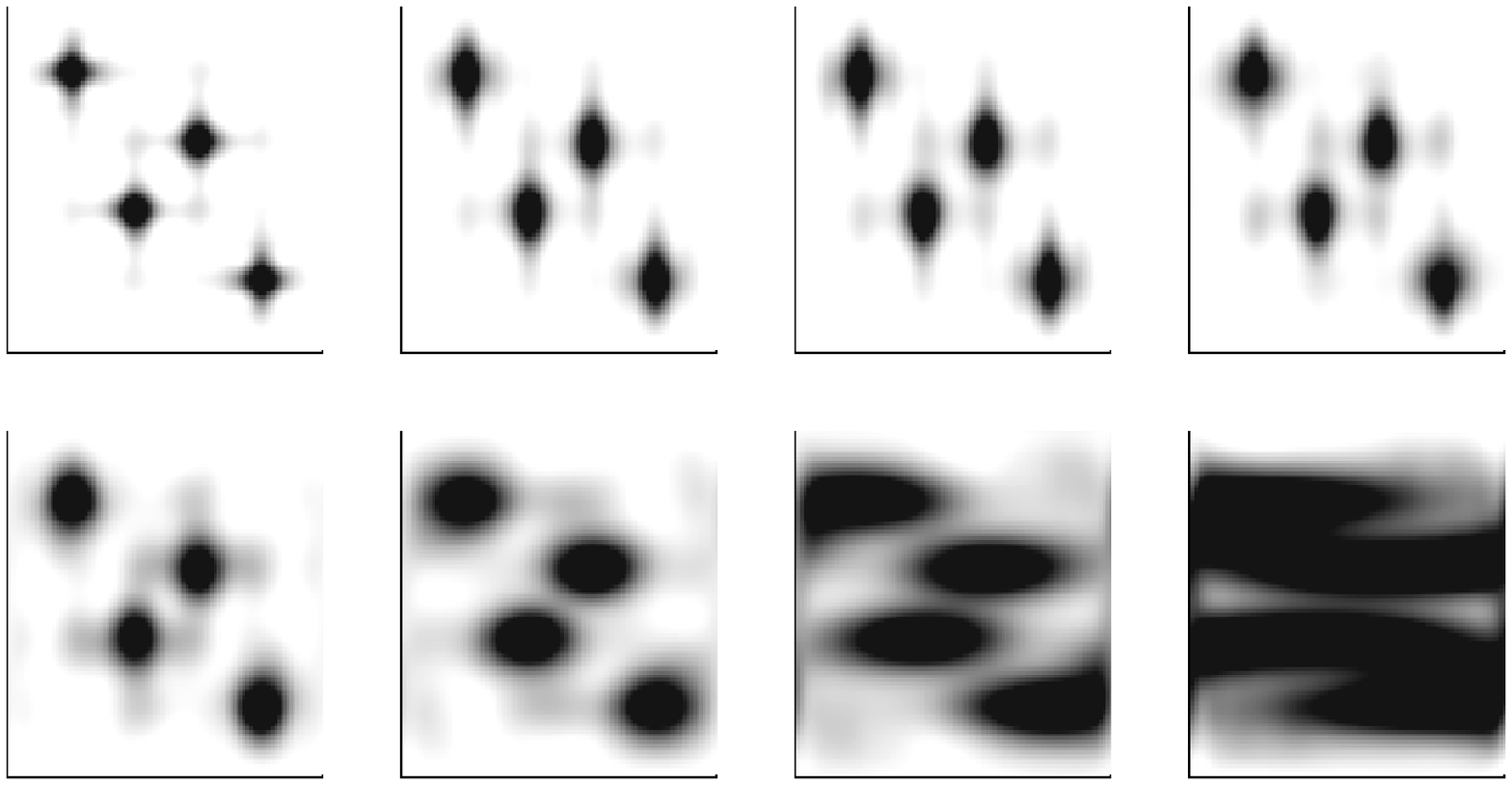}
 }

 
 \subfigure[ Map eigenstate] {
 \includegraphics[width=0.4\textwidth]{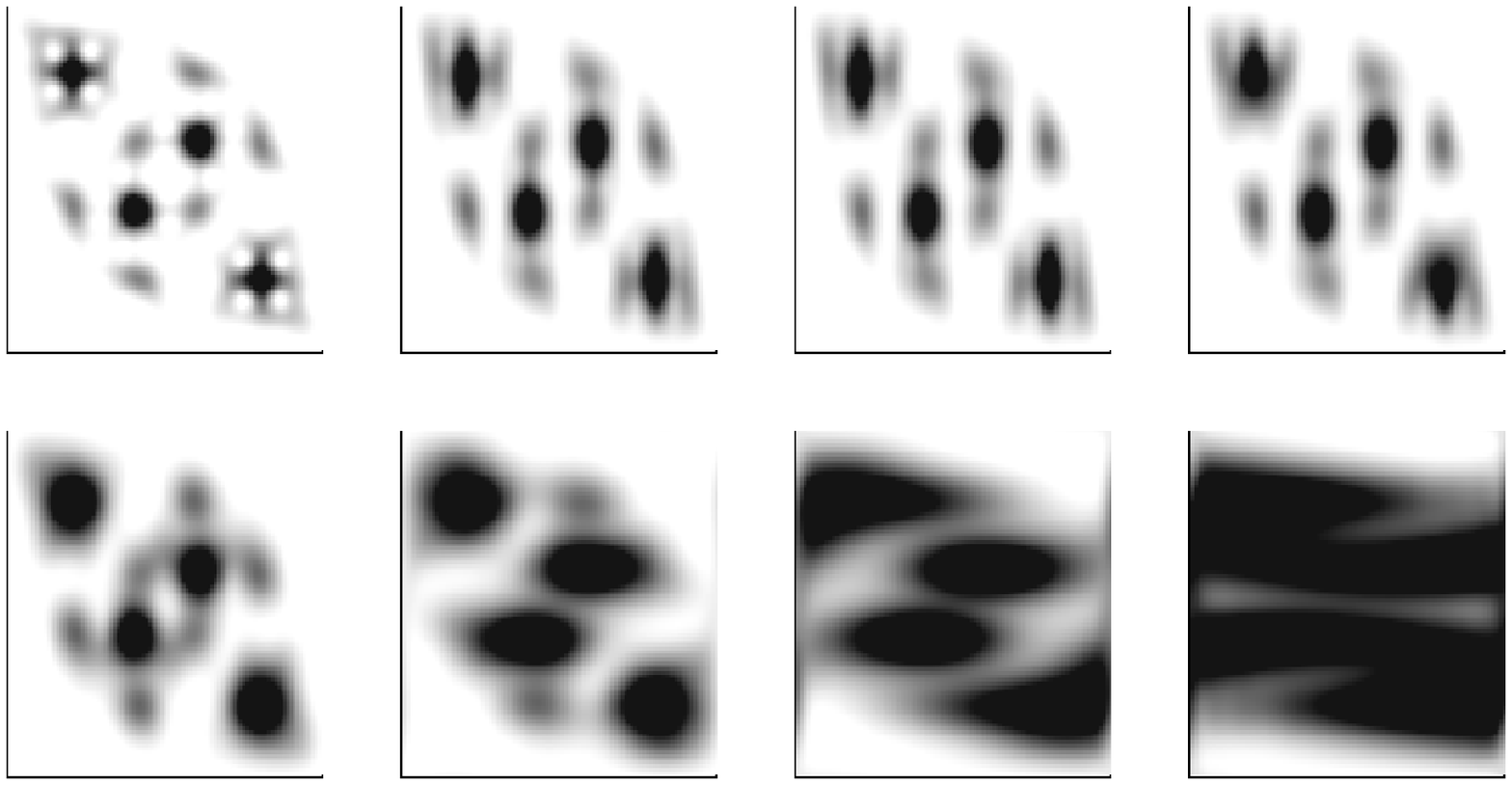}
 }

\caption{Husimi distributions for the POM and scar function associated 
to orbit $\nu=0011$ ((a) upper and (b) middle panels, respectively), 
and corresponding map eigenstate ((c) lower panel). We have used 
$\varepsilon=0.01$ and $T=1000$. Time steps go from $0$ to $7$ from left 
to right and top to bottom. The grayscale goes from 
white (minimum probability) to black (maximum probability). }
\label{HusimiLoc}
\end{figure}

We now briefly describe the scar function construction \cite{ScarFunction1,ScarFunction2,ScarFunction3}, 
the main tools we use to study the stability of classical structures. These functions are 
also essential in the semiclassical description of chaotic eigenfunctions \cite{SPOth}. 
They are wavefunctions highly localized on the stable and unstable manifolds of POs, and 
on the energy given by a Bohr-Sommerfeld quantization 
condition on the trajectory.
We are going to use a formulation suitable for maps on the torus, though 
they apply also to general flows. The first step is to
define the {\em Periodic Orbit Modes} (POMs) for maps, $|\phi_{\rm POM}^{\rm maps}\rangle$ 
as a sum of coherent states 
centered at the fixed points of a given PO $\nu$, each one having 
a phase \cite{ScarFunction3} (see Figure \ref{HusimiLoc}(a) for an example). 
Then, the scar function is obtained after applying a 
dynamical average and can be written as
\begin{equation}
\label{eq:fscarm}
|\phi_{\rm scar}^{\rm maps}
\rangle= \sum_{l=-T}^T e^{i S_{\nu} l/\hbar} \; \cos \left(\frac{\pi l}{2 T} 
\right) \; B^l \; |\phi_{\rm POM}^{\rm maps}\rangle,
\end{equation}
where $T$ stands for the number of iterations of the map up to the Ehrenfest time
$T_E=\ln{N}/\lambda$, and $S_{\nu}$ is the 
classical action of $\nu$ (see Figure \ref{HusimiLoc}(b) for an example). 
It is worth mentioning that for $T=0$ the POMs are recovered, 
but as $T$ is increased this function sharpens its quasi--energy width while extending 
in phase space along the stable and unstable manifolds of the periodic orbit. 
Eventually, it turns into a true eigenstate if the propagation time reaches the order 
of the Heisenberg time.


In our calculations we have considered five types of initial states. Besides the 
already mentioned scar states and POMs, we have studied the behavior of eigenstates 
of the map having a significant overlap with scar functions. Also, we have considered 
states developed along the stable and unstable manifolds. 
These last two cases correspond to eigenstates 
of the momentum and position operators, since 
for the baker map the direction of the manifolds coincide with those of 
the corresponding axes. In addition, we have selected $q$ and $p$ values that 
represent points belonging to the periodic orbits shown.
In summary, we compare states with relevant classical imprints 
(localized on POs and their manifolds) with strictly quantum 
states (map eigenstates). All results correspond to $N=100$.

\begin{figure}
 \includegraphics[width=0.48\textwidth]{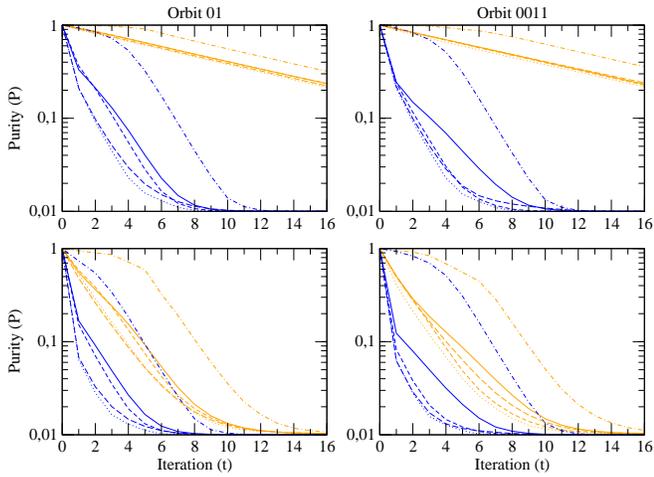} 
\caption{(Color online) Purity $P$ (logarithmic scale) as a function of time 
(measured in units of map steps). In the left column we show 
results for the orbit $\nu=01$, while in the 
right one for $\nu=0011$. Upper panels correspond to $\varepsilon=0.001$, and lower ones 
to $\varepsilon=0.01$. Orange (lighter) lines correspond to $T=1$ and blue (darker) 
lines to $T=1000$. Solid lines 
stand for scar functions, long-dashed lines for eigenstates of the map, short-dashed for 
POMs, dot-dashed for position eigenstates and dotted for momentum eigenstates. $N=100$.}
\label{Purity}
\end{figure}

The behavior of the coherence is studied through the purity, 
defined as $P=\mathrm{tr}(\rho^2)$ ($\mathrm{tr}$ denotes the trace). 
We find that the scar functions, the map eigenstates that have a maximum overlap with 
them, and the POMs, all behave in a similar way. However, the purity decay is 
faster in the case of the eigenstates when compared to that of the corresponding 
scar functions. On the other hand the position eigenstates are 
the less and the momentum eigenstates the most decoherent ones. 
We show  results for two representative orbits in the left and right columns of Figure \ref{Purity} 
(they are denoted by $\nu=01$ and $\nu=0011$, respectively). In obtaining these results we have considered 
two different couplings with the environment ($\varepsilon=0.001$ and $\varepsilon=0.01$), 
and two different values of the temperature ($T=1$ and $T=1000$).
They convalidate the hypothesis that entropy production, then entanglement 
with the environment, is governed by the classical unstable behavior of the system 
\cite{ZurekPaz}. The more localized on classically unstable regions the initial 
distribution is, the fastest its entropy production results.

\begin{figure}
\begin{center}
\includegraphics[width=0.48\textwidth]{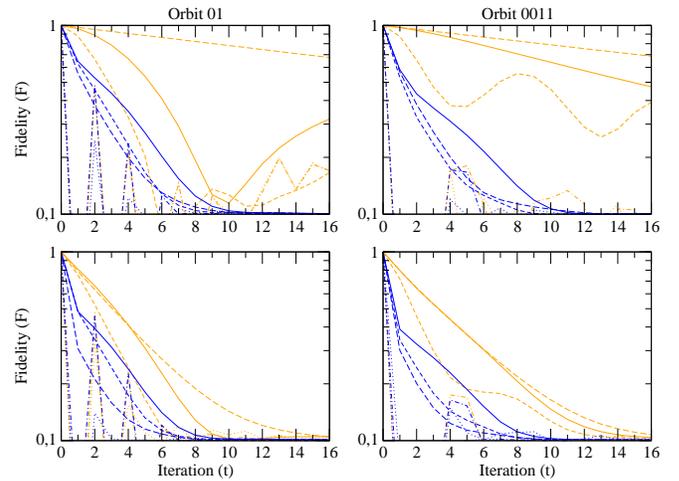} 
\caption{Fidelity $F$ (logarithmic scale) as a function of time (measured in units of map steps). 
We show results for the same orbits, use the same parameter values, and colors and patterns 
criterion as in Fig. \ref{Purity}.}
\label{Fidelity}
\end{center}
\end{figure}

In order to have a complementary stability criterion we have explored 
the fidelity calculated as $F=\sqrt{\bra{\psi} \rho \ket{\psi}}$. This 
quantity measures the degree of overlap between the pure initial state $\ket{\psi}$ and 
the evolved density matrix $\rho$. This time we find that although 
the scar functions and the corresponding map eigenstates have a 
similar behavior, the latter are more stable than the former when the 
temperature is low. For higher values the situation changes and the scar functions 
become more stable than the eigenfunctions. The coupling with the environment seems 
to play a similar role than temperature, since the decay rate difference for low values 
of $T$ shrinks as $\varepsilon$ grows. Finally, POMs fidelity decays faster, 
and the position and momentum eigenstates have the greatest 
decay rate but with big oscillations, whose periodicity is given by the dynamics. 
Hence, for low temperatures the stability is governed by the quantum mechanics of the system.
Initial states that are closer (greater overlap) to the map eigenfunctions turn out 
to be the most stable ones. In this sense, scar functions are more stable than POMs, which in turn  
are more stable than position and momentum eigenstates. When the effects of the environment 
become stronger scar functions emerge again as remarkably stable, having the lowest decay 
rate of all the initial states we have considered.
This is shown in Fig. \ref{Fidelity}, where we have used the same orbits, 
parameter values, colors and patterns as those chosen for Fig. \ref{Purity}. 

The picture is completed by means of the Husimi distributions obtained at different times. 
They can be seen in Figure \ref{HusimiLoc}, where the evolution of the POMs, the scar 
functions and the map eigenstates for the orbit $\nu=0011$ can be found. The high stability 
of the scar functions is undoubtedly demonstrated. When already the POMs shape cannot 
be distinguished anymore they still display a high level of detail, showing their 
characteristic localization along the manifolds of the corresponding orbit. Moreover, 
the eigenstate rapidly loses its original details and seems to converge to the scar function. 

In conclusion, we have verified a great degree of complementarity 
between purity and fidelity as decoherence measures. 
The loss of purity, or entropy production,
is governed by the classical instability of the system, while the fidelity 
decay for low temperatures (and weak coupling with the bath) 
is dominated by the overlap of the initial state with the map eigenfunctions.
When temperature rises the situation changes and the localization 
on classical structures plays again a fundamental role.
By analyzing these quantities we have found classical structures in quantum mechanics 
that are specially stable when subjected to environmental perturbations. 
They are the scar functions, which are associated to periodic orbits 
and the stable and unstable manifolds in their vicinity.
Husimi distributions in phase space allowed us 
to clarify this picture. We have conclusively shown that scar functions 
keep their shape virtually intact at times when the POMs have already lost 
all their characteristic features. Hence, more than just localization on the periodic 
points is needed to provide stability. Finally, by exploiting the simplicity 
of the manifolds of our map we could verify that any kind of localization on them 
is not enough to guarantee robustness against external perturbations, but the 
one provided by scar functions. 

Remarkably, the purity and fidelity loss of the map eigenstates (in this last 
case with the exception of low temperatures) is generally faster 
than that of the scar functions. Then, these latter represent the stable classical 
skeleton of the map eigenstates against environmental perturbations.
It is worth underlining that these classical structures which survive for 
longer times are also the main classical ingredient needed to construct the eigenstates, 
according to the short periodic orbit theory.

\begin{acknowledgments}

Support by CONICET is gratefully acknowledged.

\end{acknowledgments}

\end{document}